\RequirePackage[loading]{tracefnt}
\documentclass[conference]{IEEEtran}

\usepackage{comment}
\usepackage{todonotes}
\usepackage{xspace}
\usepackage[caption=false]{subfig}
\usepackage{url}
\usepackage{hyperref}
\hypersetup{ %
    colorlinks,
    linkcolor={red!40!black},
    citecolor={blue!40!black},
    urlcolor={blue!40!black}
}

\newcommand{\changed}[1]{\textcolor{black}{#1}}
\newcommand{\lm}[1]{\textcolor{black}{#1}}
\usepackage{microtype} 

\newcommand{\name}{Codehacks\xspace}
\newcommand{\hackAttempt}{1,002,339\xspace}%
\newcommand{\hacksAll}{393,382\xspace} %
\newcommand{\hacksRelevant}{288,617\xspace} %
\newcommand{\hacksOmitted}{104,765\xspace} %
\newcommand{\contestsAll}{$1,928$\xspace} %
\newcommand{\contestsRelevant}{$1,647$\xspace} %
\newcommand{\problems}{5,578\xspace} %
\newcommand{\submissions}{2,196\xspace}   %

\usepackage[style=numeric-comp, maxnames=100, abbreviate=true, dateabbrev=true, alldates=year, doi=true, isbn=false, url=false, eprint=false, language=american, giveninits=true, sortcites=true, sorting=none, backend=biber]{biblatex} %

\AtEveryBibitem{\clearlist{location}} %
\AtEveryBibitem{\clearlist{language}} %
\AtEveryBibitem{\clearfield{eprintclass}} %

\begin{document}

\thispagestyle{plain}
\pagestyle{plain}

\makeatletter
\def\ps@IEEEtitlepagestyle{%
  \def\@oddfoot{\mycopyrightnotice}%
  \def\@evenfoot{}%
}
\def\mycopyrightnotice{%
  \hspace*{3mm}\includegraphics[width=2cm]{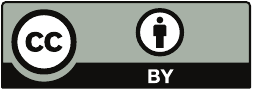}%
  \hspace*{2mm}\raisebox{2.5mm}{%
          \parbox{\columnwidth}{\footnotesize This work is licensed under a Creative Commons \\ Attribution 4.0 International (CC BY 4.0) license.}%
          \hspace*{-73pt}\mbox{\thepage}\hspace{25pt}\fbox{\parbox{.84\columnwidth}{\footnotesize\textsl{Accepted for publication at the 18th IEEE International Conference on Software Testing, Verification and Validation (ICST 2025).}}}%
  }%
  \gdef\mycopyrightnotice{}%
}
\makeatother

\title{Codehacks: A Dataset of Adversarial Tests for Competitive Programming Problems Obtained from Codeforces}

\author{\IEEEauthorblockN{Max Hort}
\IEEEauthorblockA{Simula Research Laboratory\\
Oslo, Norway\\
Email: maxh@simula.no}
\and
\IEEEauthorblockN{Leon Moonen}
\IEEEauthorblockA{Simula Research Laboratory\\
Oslo, Norway\\
Email: leon.moonen@computer.org}
}

\maketitle

\begin{abstract}
Software is used in critical applications in our day-to-day life and it is important to ensure its correctness.
One popular approach to assess correctness is to evaluate software on tests. 
If a test fails, it indicates a fault in the software under test; if all tests pass correctly, one may assume that the software is correct.
However, the reliability of these results depends on the test suite considered, 
and there is a risk of false negatives (i.e. software that passes all available tests but contains bugs because some cases are not tested).
Therefore, it is important to consider error-inducing test cases when evaluating software.

To support data-driven \lm{creation of such a test-suite}, %
\lm{which is} especially \lm{of interest for testing} software synthesized from large language models, 
we curate %
a dataset (\name) of programming problems \lm{together with corresponding} error-inducing test cases (i.e., ``hacks'').
\lm{This dataset is collected} from the wild, in particular, from the Codeforces online judge platform.
\changed{The dataset comprises \hacksRelevant hacks for \problems programming problems, each with a natural language description, as well as the source code for \submissions \lm{submitted solutions to these problems that can be broken with their corresponding hacks}.}
\end{abstract}


\IEEEpeerreviewmaketitle

\section{Introduction}

\noindent
Large Language Models (LLMs) are increasingly being used to support us in our daily lives and have achieved high competitive performance on a variety of software engineering tasks (e.g., bug fixing, defect detection, program synthesis, program translation)~\cite{niu2022:deep}. 
However, LLMs have been shown to \lm{`hallucinate' or `confabulate',} generating responses that may appear plausible but are incorrect.
Although this has been \lm{mostly discussed in the context of natural language chats,}
it can also cause problems for software engineering tasks, such as code synthesis~\cite{fan2023:large}. 
Therefore, it is of great importance to verify and ensure the correctness of the synthesized code~\cite{le2019:reliability}. 
One approach frequently used in practice is execution-guided evaluation~\cite{chen2022:codet}, 
which applies a pre-defined test suite (e.g., unit tests, input-output pairs) to the generated code. 
If one of the tests fails, the snippet is treated as incorrect. 
If all tests pass, it is treated as a correct program synthesized by an LLM.

\lm{However, tests and, in particular, failure-inducing tests, are expensive and time-consuming to create~\cite{schafer2023:adaptive,chen2022:codet,li2023:finding}.
As a result, there is a risk of false negatives where the code passes all available tests but still contains bugs because some cases were overlooked.
To find such false negatives and ensure correctness of the code, additional tests are needed.}

An \textbf{unexploited resource} in this regard is the online judge platform \emph{Codeforces}.
Online judges are platforms that allow users to participate in programming competitions and solve programming tasks, often in the programming language of their choice.
User submissions are evaluated on predefined test suites, and \lm{much care is taken to ensure that} the test suites are comprehensive~\cite{liu2023:who}.
However, it has recently been shown that these test suites do not always cover all cases, allowing false negative submissions to slip through~\cite{liu2023:your,liu2023:who}.

\begin{figure}
\centering
\begin{minipage}{\linewidth}
\centering
 \subfloat[Problem description.]{%
  \includegraphics[width=0.76\columnwidth,trim={0mm 7mm 0mm 0mm},clip]{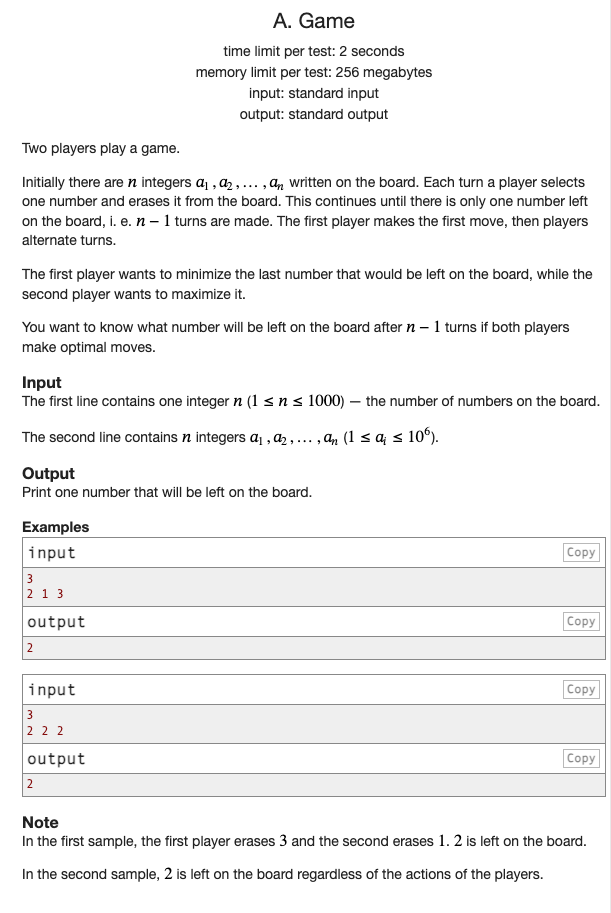}
}
\end{minipage}
\begin{minipage}{\linewidth}
\centering
 \subfloat[Accepted solution.]{%
  \includegraphics[width=0.38\columnwidth]{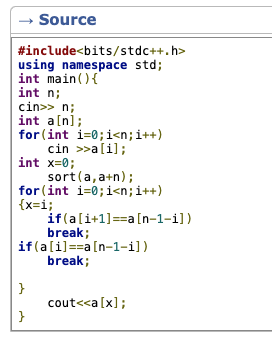}
}
\hspace{7mm}
\subfloat[Successful hacking attempt.]{%
  \includegraphics[width=0.38\columnwidth,trim={0mm 4mm 0mm 4mm},clip]{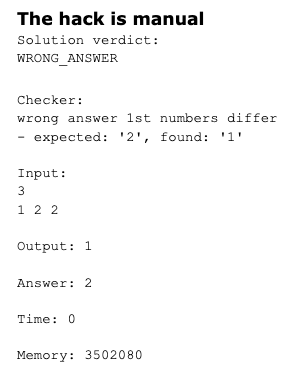}
}
\end{minipage}

  \caption{Example of a hacked Codeforces submission with corresponding problem description and hacking attempt.}
  \label{figure:example}
\end{figure}

Unlike other online judge platforms, Codeforces provides competitors the opportunity to identify such false negative submissions during a competition as a means to increase their score. 
They can do this by ``hacking'' the submissions of other competitors that already passed the predefined test suite. 
The submission of user \textit{A} is hacked by user \textit{B}, if \textit{B} can find an input for which the submission fails (e.g., it generates a different output than a pre-specified solution). 
Figure \ref{figure:example} illustrates an example hack obtained from a Codeforces contest.
It should be noted that an unsuccessful hacking attempt by user \textit{B} results in a penalty to their score.
The successful hacks from Codeforces provide a valuable \lm{learning} resource to support the \lm{data-driven} creation of test inputs to find false negative submissions and evaluate the quality of synthesized code.

This paper introduces \emph{\name,} a \textbf{novel dataset} curating failure-inducing test cases based on ``hacks'' that are automatically collected from the Codeforces coding platform.
As online judges already evaluate submissions with a multitude of tests, these represent edge cases that are costly to create manually and are valuable resources for future test generation and validation approaches for synthesized code, in particular by LLMs.
\lm{At the initial release, the dataset comprises of \hacksRelevant hacks for \problems programming problems, each with a natural language description, as well as the source code for \submissions submitted solutions to these problems that can be broken with their corresponding hacks.}
The necessary resources (dataset and scripts) to update, replicate and build on our work are provided at: \textbf{\url{https://doi.org/10.6084/m9.figshare.24773754}}

\section{Related Work}

\subsection{Online Judge Datasets}
\noindent
Online judges provide programming problems to a wide audience and allow them to submit their own solutions which are evaluated on test suites to verify their correctness. The public sharing of this data allowed for the creation of multiple datasets that provide ample resources to support software engineering tasks.
Over the last few years, datasets have been created that utilize data from online judge platforms such as Codewars, AtCoder, Kattis, Codeforces, Google Code Jam, CodeChef, and Hackerearth~\cite{hendrycks2021:measuring,ullah2019:cyber,li2022:competitionlevel}.

One of the largest and most frequently used data sets is CodeNet by Puri et al.~\cite{puri2021:codenet}. 
CodeNet consists of over 14 million code samples in 55 programming languages, with sample input and output pairs (tests) for 98.5\% of the code. These data are crawled from \lm{the Aizu and AtCoder online judge platforms}.
Subsets of CodeNet have been used to create program repair datasets~\cite{prenner2023:runbugrun}, 
\lm{as well as incorporated in CodeContests, the dataset created for training the AlphaCode language model~\cite{li2022:competitionlevel}.
The complete CodeContests dataset contains problem descriptions, submissions, and test cases 
that were collected from the Aizu, AtCoder, CodeChef, Codeforces, 
and HackerEarth platforms.\footnote{~\url{https://github.com/deepmind/code_contests}}
The collected submissions are written in the three most frequently used programming languages: Python, Java, and C++.
}

\lm{To evaluate whether the tests for AtCoder are effective in detecting false negatives, Liu et al.~\cite{liu2023:who} extracted 541,552 accepted solutions from 939 coding problems from the CodeContests dataset. 
After randomly generating additional tests, they found false negative submissions for 43.1\% of the problems.
A subset of these data containing 3,043 false negative submissions is shared as ``TrickyBugs'' to stimulate future research~\cite{liu2024:trickybugs}. }

\lm{In addition}, there are datasets that are completely based on Codeforces problems.
An early dataset created in 2017 by Tan et al.~\cite{tan2017:codeflaws} is called CodeFlaws. CodeFlaws is designed to support program repair tasks and consists of 7,436 programs from Codeforces (i.e., a user submission to a specific problem). Each submission consists of a rejected submission, classified in one of 39 defect classes, and an accepted one.  %
Code4Bench introduced a richer dataset with 3,421,357 Codeforces programs written in 28 programming languages~\cite{majd2019:code4bench}.

However, none of the existing works made use of the ``hacking'' functionality that Codeforces provides to curate their datasets.
In particular, unlike the randomly generated tests by Liu et al.~\cite{liu2023:who}, these hacks are submitted by human contestants, which is helpful when evaluating code synthesis tools, trained on human-written code.

\subsection{Program Synthesis}
\noindent
Program synthesis aims at the generation of a program with a specified target language, based on natural language descriptions or specifications of input-output pairs~\cite{chen2019:executionguided}.
Advances in LLMs for code synthesis introduced well-performing models, such as AlphaCode which was able to rank in the top 54.3\% in Codeforces coding competitions~\cite{li2022:competitionlevel}.
AlphaCode is a transformer model pre-trained on source code from GitHub repositories and \lm{fine-tuned on the CodeContests dataset}.
Another LLM for code generation is Codex~\cite{chen2021:evaluating}, a GPT model fine-tuned on GitHub repositories for writing Python code.
 
Jain et al.~\cite{jain2022:jigsaw} proposed an interactive system for code synthesis called Jigsaw.
Jigsaw allows users to describe the program they want to generate with natural language and test cases (input and output pairs). 
Generation is then carried out by integrated LLMs, such as GPT-3 and Codex.

\lm{While Li et al.~\cite{li2022:competitionlevel} evaluate AlphaCode on Codeforces contests, they mention that hacking was not performed during their evaluation of AlphaCode} (i.e., the solutions submitted by AlphaCode to the contests cannot be hacked unlike regular submissions).
The use of hacks as an additional resource for training was not considered.

\subsection{Test Generation}
\noindent
Tests are required to ascertain the correctness of software. 
Given the effort, cost, and time required to generate suitable tests for source code, 
test generation techniques have been developed to support software engineers with this task. 
Here, we outline recent approaches for test generation that utilize LLMs and could benefit from our \emph{\name} dataset.

Tufano et al.~\cite{tufano2021:unit} proposed ATHENATEST which treats unit test generation as a sequence-to-sequence learning task.
Schaefer et al.~\cite{schafer2023:adaptive} introduced TESTPILOT, a unit test generator based on Codex~\cite{chen2021:evaluating}. Codex is utilized without further retraining, solely by prompts including the respective function and examples. 

Liu et al.~\cite{liu2023:your} proposed a benchmarking framework (EvalPlus) to assess the correctness of LLM-synthesized code.
EvalPlus employs both LLMs and mutation-based strategies to generate additional tests.
In particular, ChatGPT is used to generate seed tests that are modified by mutation strategies.
The prompt for ChatGPT includes the program solution, example test inputs, as well as an encouragement to generate interesting test inputs.

\lm{In many studies, the LLMs are used \emph{off-the-shelf}, and the investigation is aimed at finding suitable prompts for the generation of tests~\cite{chen2022:codet}.
Our work is orthogonal to such studies, as we aim to curate a high-quality dataset that can then be used to fine-tune an LLM for the task of generating failure-inducing tests given a natural language description of a programming problem.
}

\section{Codeforces Dataset - Collection and Curation}
\noindent
Codeforces is a platform designed for practicing and participating in programming contests~\cite{rahman2023:predicting,majd2019:code4bench}.
Codeforces hosts contests consisting of multiple problems. Users can join a contest live, during which they can work on the problems for a designated time duration, but they can also solve problems offline, after the contests are over.
Competitors have a free choice of programming language, while test inputs are independent of the programming language used~\cite{majd2019:code4bench}.
During the duration of the contest, users can hack the submission of other contestants to gain additional points and improve their ranking. 
Hereby, an incorrect hacking attempt causes a point deduction. 
In addition, there are educational contests that allow for hacking up to 12 hours after contest completion for learning purposes.\footnote{~\url{https://codeforces.com/blog/entry/107753}}

In the following, we outline the steps that were performed to collect hacks \changed{and submissions} from competitions held on Codeforces and create the \name dataset (Section~\ref{data:collection}). Section~\ref{data:verdicts} provides details on the collected hacking attempts, as well as a discussion of our choices on which hacks to include in the dataset. Section~\ref{data:types} describes the programming problems for which hacks were collected\lm{, and Section~\ref{data:limitations} discusses possible limitations of the collection process}. 

\subsection{Data Collection}
\label{data:collection}
\noindent
To collect relevant data from Codeforces, we use their API\footnote{~\url{https://codeforces.com/apiHelp}} and a publicly available crawler.\footnote{~\url{https://github.com/Nymphet/codeforces-crawler/tree/master}\label{foot:crawler}}
With help of the Codeforces API, we obtain information about hacks. In particular, we first used the API to find the IDs for every contest held on Codeforces and afterwards obtain a list of hacking attempts for each contest.
In total, from the \contestsAll contests currently on Codeforces, \contestsRelevant contests have hacking attempts from users, resulting in \hacksAll successful hacks.\footnote{~We accessed the API on the 18th of November 2024.}

\begin{figure}
\centering
  \includegraphics[width=\columnwidth]{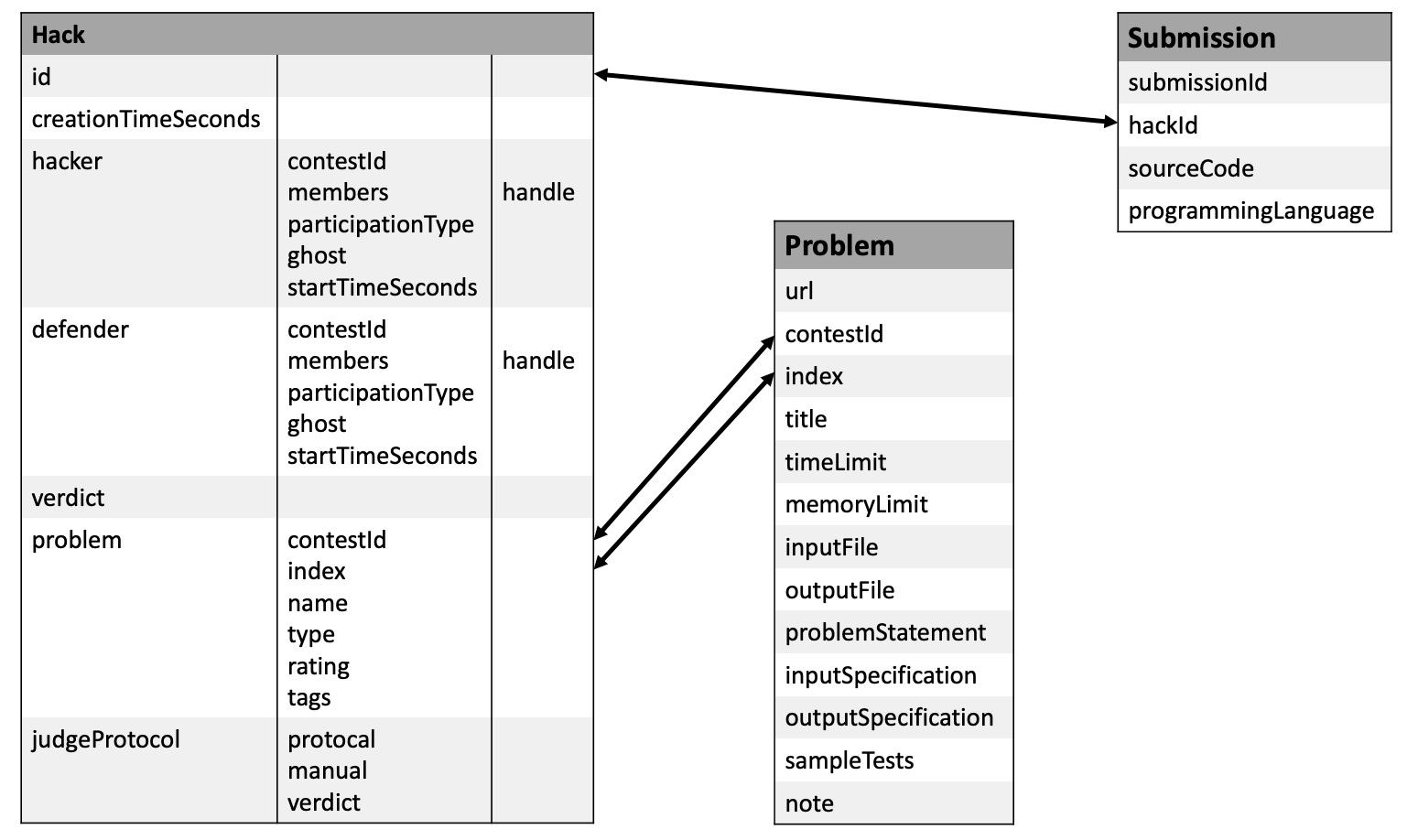}
  \caption{Structure of the collected dataset.}
  \label{fig:dataset}
\end{figure}

While the API call provides useful information, such as the input used for a hacking attempt, the problem description is not provided.
Therefore, we follow Tan et al.~\cite{tan2017:codeflaws}, to use and modify a publicly available crawler\footref{foot:crawler} to extract problem descriptions for each of the \problems problems with successful hacks.
A description of the information obtained for hacks and problems is shown in Figure~\ref{fig:dataset}.

\changed{
Continuing the crawling process to gather further details on submissions, such as their source code, turned out infeasible as they are part of the ``robots.txt''.
Therefore, we used Code4Bench~\cite{majd2019:code4bench}, a dataset from 2018 with $3,421,357$ submissions, to match the submission for our hacks with the already collected dataset.
In total, the hacks from our dataset matched with \submissions submissions from Code4Bench.
We add the respective source code and programming language used to our dataset, as outlined in Figure~\ref{fig:dataset}.}

According to the Codeforces data sharing guidelines, we provide the URL to each problem description crawled.\footnote{~\url{https://codeforces.com/blog/entry/967}}
An explanation of the required steps and relevant source code is provided in our online Appendix.\footnote{~\url{https://doi.org/10.6084/m9.figshare.24773754}}

\begin{comment}
Overall, we extracted the following information for each problem:
\begin{itemize}
    \item Url (p\_url)
    \item Contest ID (p\_contest)
    \item Problem ID (p\_index)
    \item Title (p\_title)
    \item Time limit (p\_time\_limit)
    \item Memory Limit (p\_memory\_limit)
    \item Input file type (p\_input\_file)
    \item Output file type (p\_output\_file) 
    \item Problem statement (p\_statement)
    \item Input specification (p\_input\_specification)
    \item Output specification (p\_output\_specification) / Interaction
    \item Sample tests (p\_sample\_tests)
    \item Note (p\_note)
\end{itemize}
\end{comment}

\subsection{Hack Verdicts}
\label{data:verdicts}

\begin{figure}
\centering
  \includegraphics[width=.45\textwidth,trim={20mm 20mm 20mm 0mm}, clip]{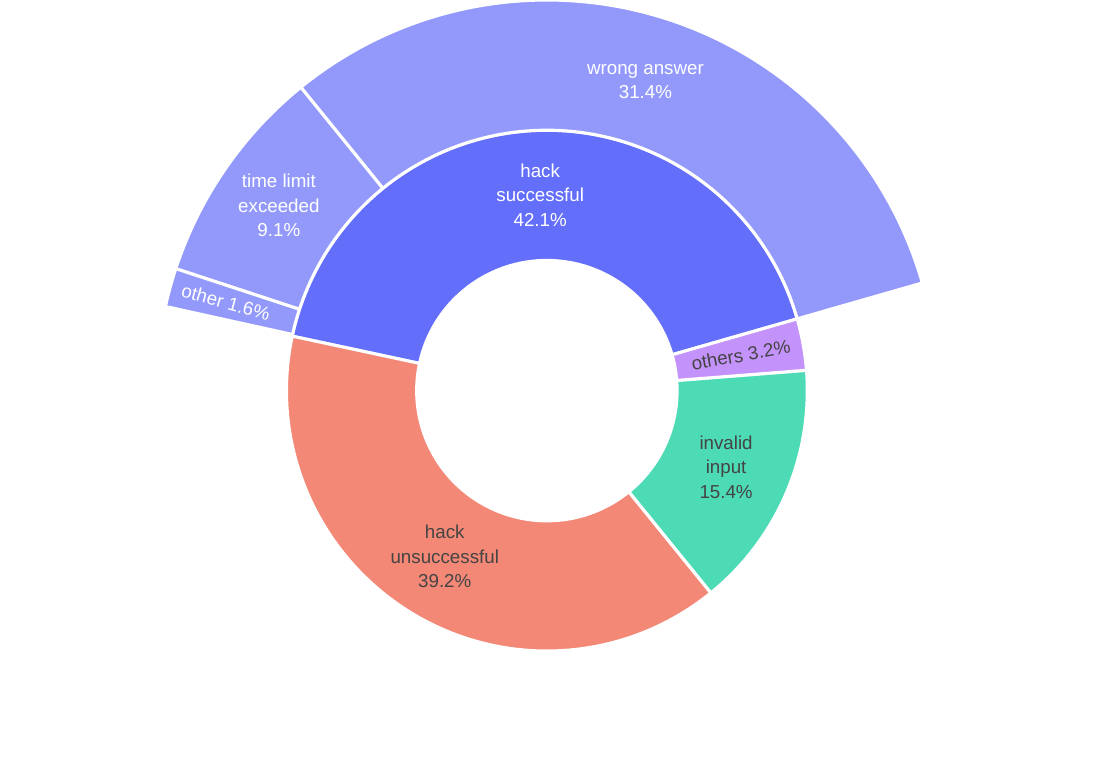}
  \caption{Verdicts of the submitted hacks.}
  \label{fig:verdicts}
\end{figure}

Each hacking attempt is evaluated by the Codeforces platform to determine whether it was successful or not (i.e., whether the hacker is able to show erroneous behavior of an accepted submission).
For example, to make a submission fail due to a ``time limit exceeded'' error, one often needs to create as many inputs lines as allowed by the problem specification.
To make a submission fail due to a wrong answer, one needs to find an input that is not covered by the online judges test set. 
Figure~\ref{fig:verdicts} illustrates the distribution of the verdicts among the \hackAttempt hacking attempts, of which \hacksAll (42.1\%) are successful.
The majority of successful attempts force the submission to generate a wrong answer (31.4\% of total) and the next most successful type group makes the submission exceed predefined time limits (9.1\% of total). 
The remaining types of errors that are induced include runtime errors and exceeding memory limits.
\changed{39.2\% of the submitted hacks are not able to point out errors in the submissions (hack unsuccessful), 
while 15.4\% do not conform to the input specifications of the problems and are therefore not executed (invalid input).
We also observed that 3.2\% of the hacking attempts received other tags (i.e., generator incompilable, generator crashed, ignored, other, testing).
\lm{We treat these, as well as invalid inputs, as unsuccessful hacking attempts and exclude them from the dataset}. 
}

We note that the Codeforces API abbreviates long input sequences (e.g., an input can consist of more than 10,000 lines) by substituting parts of the input with ``...''~\cite{majd2019:code4bench}. 
This is the case for \hacksOmitted of \hacksAll successful hacking attempts.
These are omitted from the collection as their input cannot be reproduced, leaving \hacksRelevant hacks for the dataset collection.

\begin{comment}
\begin{figure}[tb]
\centering
  \includegraphics[width=\columnwidth]{images/verdicts2.pdf}
  \caption{Verdicts of the submitted hacks.2}
  \label{fig:verdicts}
\end{figure}
\end{comment}

\subsection{Problem Types}
\label{data:types}
\begin{figure}
\centering
  \includegraphics[width=\columnwidth]{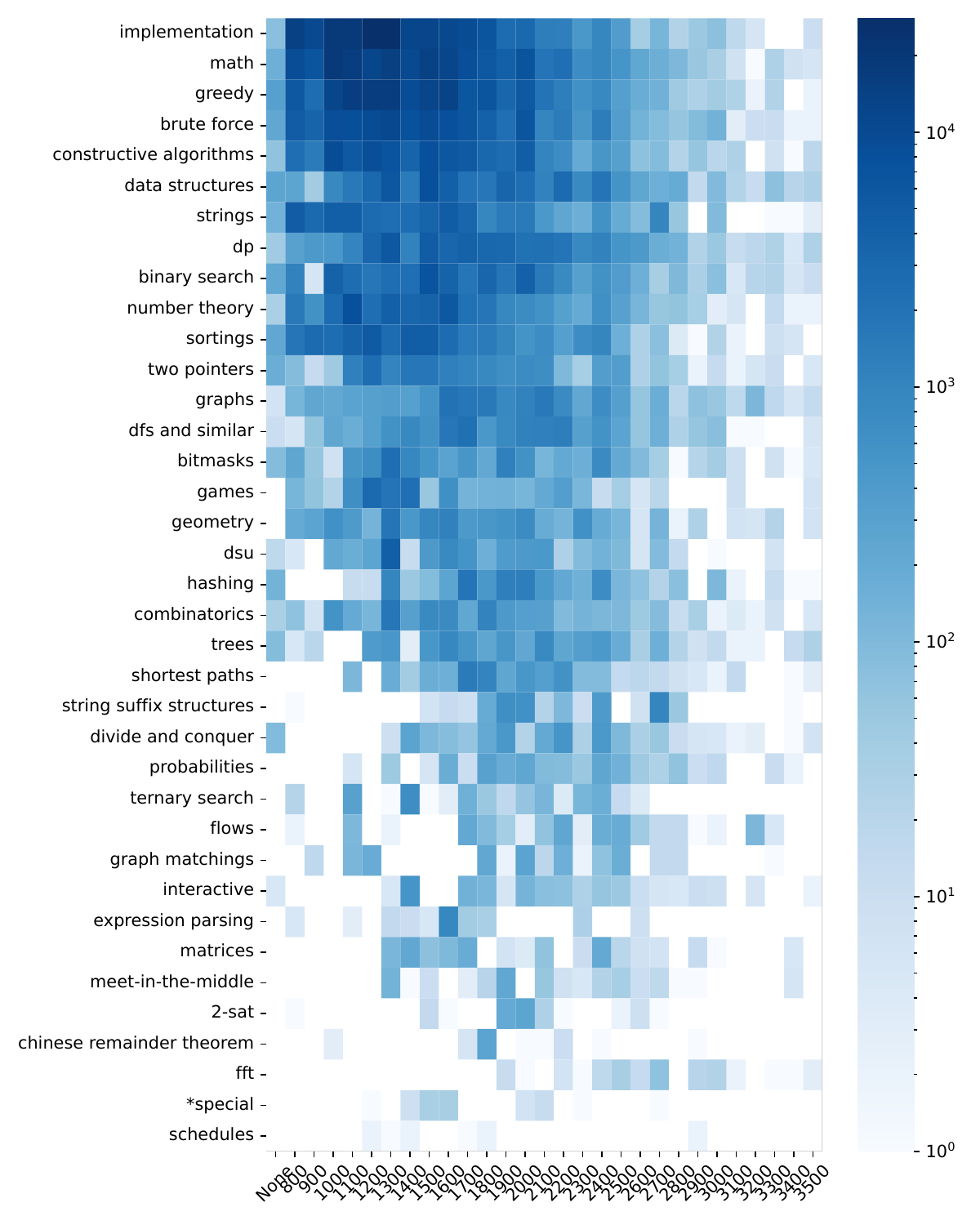}
  \caption{Distribution of hacks per problem tag (y-axis) and difficulty (x-axis).}
  \label{fig:heatmap}
\end{figure}

\noindent
Each Codeforces problem is categorized based on different tags that indicate the type of problem (e.g., dynamic programming). Moreover, problems have a certain difficulty level ranging from 800 to 3500, to indicate how difficult it is for a user to solve this problem.\footnote{~Note that 2,415 of the problems did not have a specified difficulty level.} 
\lm{In Figure \ref{fig:heatmap}, we illustrate frequency of problem types and their respective difficulty level for the collected hacks.}
The tags are ordered by frequency in the datasets, showing that the most frequently used tags to describe problems with successful hacks are: implementation, math, greedy.
Moreover, we can observe that there is a larger number of hacks for problems with lower difficulty levels, which could be caused by the overall distribution of problems on Codeforces (e.g., there are more problems with lower difficulty), or the proficiency level of contestants (e.g., contestants that solve problems with low difficulty levels have a lower rating and therefore create solutions that are ``hackable'').

\begin{comment}
\begin{figure}
\centering
  \includegraphics[width=\columnwidth]{images/heatmap.pdf}
  \caption{Distribution of hacks per problem tag (y-axis) and difficulty (x-axis).}
  \label{fig:heatmap}
\end{figure}
\end{comment}

\subsection{Limitations}
\label{data:limitations}

\noindent
The quality of the successful hacks depends on the pre-defined test suite employed by Codeforces for each of the problems.
There are no regulations on how many tests a problem should include, and therefore it is not possible to judge the coverage or completeness of the test suites.
This could lead to some of the hacks collected being ``easy'' due to small initial test suites, while other problems \lm{with more extensive test suites may not be covered because no successful hacks were submitted for them}.

\section{Conclusions and Future Usage}
\noindent
\lm{This paper introduces \name, a dataset of user-submitted hacks that reveal errors in submissions to programming problems that the standard test suites on Codeforces are not able to cover.}
\changed{In total, this data set comprises \hacksRelevant successful hacking attempts and \submissions source code submissions for \problems programming problems. For each of the problems, we include the natural language description.}

Curating a collection of failure-inducing test cases for programming problems is of interest,
as such test cases can be difficult and costly to find. 
We hope  that the provided dataset can function as a valuable resource for software testing and test generation.

\lm{One promising application area is to improve code synthesis.
It is known that high-quality tests can help improve code synthesis~\cite{austin2021:program}, 
and that it is important to consider edge cases when testing programs, as these are the most likely to point out flaws in the program logic.
Our dataset can be used to fine-tune an LLM for generating such edge cases (i.e., generate likely error-inducing tests),
based on natural language descriptions of the problem for which code is synthesized.}

Another promising field for applying our dataset is test generation \lm{in an adversarial setting}.
In this case, an LLM such as AlphaCode or Codex is used to synthesize code from natural language descriptions, 
while another model is trained to \lm{generate} challenging test inputs from natural language descriptions. 
These test inputs are then used to determine the quality of \lm{the code generated by} the first model.
\lm{Similarly to the iterative approach of Liventsev et al.~\cite{liventsev2023:fully}, 
who generated code and utilized feedback from failed tests for iterative debugging and repair, 
this adversarial framework can be extended to iteratively generate novel challenging tests, forcing the code synthesizer to generate more robust code.
}

\section*{Acknowledgment}
\noindent
We would like to thank \href{https://codeforces.com/profile/MikeMirzayanov}{\textbf{MikeMirzayanov}} for creating the Polygon and Codeforces platforms.

This work is supported by the Research Council of Norway through the secureIT project (IKTPLUSS \#288787), and by the European Union through the Horizon Europe Marie Sk\l{}odowska-Curie Actions (\#101151798).

\printbibliography

\end{document}